\documentstyle[aps,prl,floats,epsfig]{revtex}
\begin{document}
\draft
\twocolumn[\hsize\textwidth\columnwidth\hsize\csname @twocolumnfalse\endcsname

\title{On the nature of the depinning transition in type-II superconductors}
\author{N. L\"utke-Entrup, B. Pla\c{c}ais, P. Mathieu and Y. Simon}
\address{ Laboratoire de Physique de la Mati\`ere 
Condens\'ee de l'Ecole Normale Sup\'erieure\\CNRS URA 1437, 24 rue Lhomond, F-75231 
Paris Cedex5}

\date{\today} \maketitle

\begin{abstract}
The surface impedance $Z(f)$ of conventional isotropic materials has 
been carefully measured for frequencies $f$ ranging from 1\thinspace kHz to 
3\thinspace MHz, allowing a detailed investigation of the depinning transition.
Our results exhibit the irrelevance of classical ideas on the dynamics 
of vortex pinning. 
We propose a new picture, where the linear ac response is entirely 
governed by disordered boundary conditions of a rough surface, 
whereas in the bulk vortices respond freely.  
The universal law for $Z(f)$ thus predicted is in remarkable agreement with 
experiment, and tentatively applies to microwave data in YBaCuO films. 
\end{abstract}
\pacs{PACS numbers:  74.60.Ge, 74.25.Nf. }
]

A perfect sample of a type-II superconductor in the vortex (or 
mixed) state would be transparent to an electromagnetic wave 
at very low frequencies.  
But defects are always present and strongly alter the quasistatic 
and  low-frequency response; 
low frequencies here means $\Omega=2\pi f\ll \Omega_d$, 
a so-called ``depinning frequency'' \cite{Gittleman66} depending on the
material and vortex density.  
It is important for applications to know 
what kind of defects can pin vortices, how they hinder small vortex 
oscillations and thereby restrain the penetration of an ac ripple.  In 
this respect, a study at low levels of excitation of both the resistive and 
inductive part of the surface impedance $Z(\Omega)=R-iX$ as a function 
of the frequency provides much information about the dynamics of pinning.  
It is generally accepted that bulk pinning centers limit the quasistatic skin 
effect to a pinning (or Campbell's) length $\lambda_C\sim1-100\,\mu$m, 
while dissipation is vanishingly small, as observed \cite{Gittleman66,Campbell69}.
No model however has been able to account for variations of $Z$ at higher frequencies.  
In particular, as the first increasing of $R(f)$ is stronger than expected, 
the understanding of dissipation remains a puzzling problem, 
including in high $T_c$ materials \cite{Belk96}.  

\begin{figure}[!!!t]
\centerline{\epsfig{file=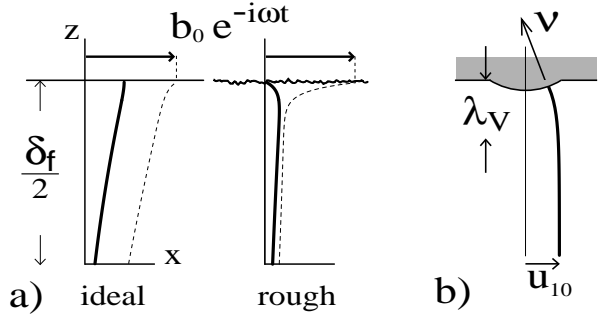, scale=0.3}}
\caption{ {\bf a)} Vortex lines $u(z)$ (full lines), and field 
profiles $b(z)$ (dashed lines) near one face of a thick slab are 
sketched with arbitrary units; for clarity, the actual length scaling, 
$u\ll a\ll\delta_f$, is not preserved.  For an ideal surface,
vortices end normal to the plane boundary, the 
weight of the $k_2$-mode {\protect\cite{Placais96}} is lowered, and a large 
normal-like skin effect is observed.  For a real rough surface, 
a vortex-slippage effect at the surface induces a relatively strong 
bending of vortices over the depth $\lambda_V\sim a$, so that 
non-dissipative currents associated with this vortex curvature 
{\protect\cite{Mathieu88,Hocquet92}} are greatly screening the 
exciting field.  {\bf b)} An isolated superfluid vortex in a rotating 
box of helium II terminates at a wall asperity.  
If it is acted on in the bulk, the vortex 
line bends near the wall so as to keep on ending normal to the 
surface, making thus an angle with the mean smooth surface.  }
\label{vrock}
\end{figure}

Experiments are performed on a series of slabs of cold-rolled 
polycrystalline PbIn and pure single-crystralline Nb.  The slabs 
($xy$) are immersed in a normal magnetic field $\bbox{B}$; unless 
specified their thickness $2d$ is much larger than the flux-flow 
penetration depth $\delta_f$ (see below).  At equilibrium, up to the 
upper critical field $B_{c2}$, a regular lattice of vortex lines 
parallel to $z$ is formed, with the density $n=B/\varphi_0$, where 
$\varphi_0$ is the flux quantum.  Both faces of the slab, $z=\pm d$, 
are then subjected to an ac magnetic field $b_0 e^{-i\Omega t}$ 
parallel to the length ($x$-direction) of the sample.  Under such 
conditions, induced currents and electric fields, $J(z)$ and $e(z) 
e^{-i\Omega t}$, are along the $y$-direction, while vortices oscillate 
in the $xz$-planes.  For low exciting fields ($b_0\sim 1\,\mu$T), 
vortex displacements $u(z)\!\sim\!1\,$\AA\ are very small compared 
with the vortex spacing $a\simeq n^{-\frac{1}{2}}$ ($\sim 1000\,$\AA, 
for $B\sim 0.1\,$T) (Fig.~\ref{vrock}a).  The electric field $e_0$ at 
the surface $z=d$, $e_0=e(d)=-e(-d)$, is measured by means of a 
pick-up wound coil.  The main experimental difficulty in such 
measurements is to ensure a precise calibration of the phase $\varphi$ 
of $e_0$ (within better than $0.5^\circ$ at 100 kHz).  Thus we get the 
surface impedance of the slab, defined as the ratio $\mu_0 e_0/b_0$.  
Putting
\begin{equation}
\frac{iZ}{\mu_0\Omega}=\frac{ie_0}{\Omega b_0}=\lambda^*= 
\lambda^\prime \!+\! i \lambda^{\prime\prime}= \Lambda 
e^{i\varphi}\qquad,
\label{zede}\end{equation}
the ac response will be conveniently expressed in terms of the complex penetration 
length $\lambda^*$ \cite{Coffey92}.  
As easily seen, $2b_0\Lambda$ (a factor 2 
for two faces) represents the amplitude of the ac magnetic flux penetrating 
the slab per unit length along $y$.  
The length $\lambda''$ measures the dissipation, as $\lambda''/\Lambda=R/|Z|$ 
is the sine of the loss angle $\varphi$.

The analysis of the ideal response, though it is not observed 
(unless $\Omega\gg\Omega_d$), is an important step in our argument.
A perfect slab would behave like a linear continuous medium, 
of resistivity $\rho_f$ and  permeability $\mu=\mu_r\mu_0$ ($0 < \mu_r < 1$). 
Here $\rho_f\simeq\rho_n B/B_{c_2}$ is the flux-flow resistivity, 
and $\mu$ is the effective ``diamagnetic permeability'' of the mixed 
state \cite{Sonin92,Vasseur97};
$\mu_r$ increases steadily with the vortex density and rapidly approaches unity 
(typically $\mu_r>0.9$ for $B\gtrsim 0.2 B_{c_2}$).  
In the absence of pinning, an electromagnetic wave, 
$b\propto e^{\pm ik_1z} e^{-i\Omega t}$, $e =\mp\Omega b/k_1$, 
can propagate in the bulk, according to the simple equation of dispersion 
$k_1^2=i\mu\Omega/\rho_f=2i/\delta_f^2$  \cite{Coffey92}.  
The wave field $b(z)$ would be accordingly: 
$b_{10}\cosh (ik_1 z) / \cosh (ik_1 d)$, 
where $b_{10}=\mu_r b_0$ if one makes allowance for a surface screening by 
diamagnetic currents. 
This leads to $\lambda^*_{\rm ideal}=\mu_r \lambda_f \tanh (d/\lambda_f)$,
where $\lambda_f=(1+i)\delta_f/2$. 
Assuming $\mu_r\simeq 1$, 
this (undisputed) result involves all features of a normal skin effect.  
In the so-defined \emph{thick limit} (thick slabs and/or high frequencies), 
say $d\gtrsim 2\delta_f$, $\lambda^*_{\rm ideal} = \mu_r\lambda_f \simeq \lambda_f$, 
so that $\lambda'=\lambda''=\delta_f/2$.  
In the \emph{thin limit}, say $d\lesssim\delta_f$, 
$\lambda^*_{\rm ideal} = \mu_r d \simeq d$, 
which means perfect transparency.

\begin{figure}[!!!t]
\centerline{\epsfig{file=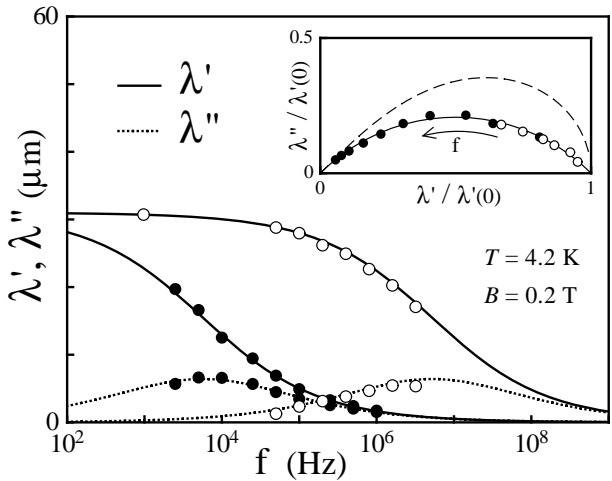, scale=0.45}}
\caption{The frequency dependence of the effective penetration depth 
$\lambda^*= \lambda'+i\lambda''$ in the thick limit 
($d\,\,{\protect\gtrsim}\,\,2\delta_f$).  Experimental data: 
({\Large$\circ$}) Pb$_{.82}$In$_{.18}$ ($2d=1.26\,$mm, 
$\rho_f=4.8\,\mu\Omega$.cm, $\Omega_d/2\pi=6\,$MHz).  
({\Large$\bullet$}) pure niobium ($2d=0.85\,$mm, 
$\rho_f=4.3\,\mbox{n}\Omega$.cm, $\Omega_d/2\pi=6\,$kHz).  Full lines 
are theoretical fits using Eq.(\ref{MSP}), where $\lambda_S$ is the 
only adjustable parameter; the flux-flow resistivity $\rho_f$ (then 
$\delta_f$ or $\lambda_f$) is measured from the dc voltage-current 
characteristics.  According to Eq.(\ref{MSP}) the universal Argand 
diagram of $\lambda^*(f)/\lambda^*(0)$ is the quarter circle shown in 
the inset: $\left[1+(\Omega/\Omega_d)e^{-i\pi/4}\right]^{-1}$.  For 
comparison the dashed line is the diagram predicted by the Campbell 
equation (\ref{CCC}).}
\label{lambda}
\end{figure}

\begin{figure}[!!!t]
\centerline{\epsfig{file=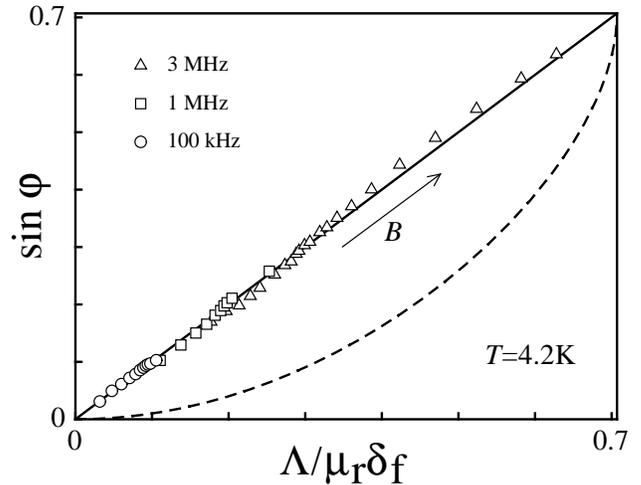, scale=0.45}}
\caption{ The complex penetration depth $\lambda^*=\Lambda 
e^{i\varphi}$ of the PbIn slab referred to in Fig.  \ref{lambda} has 
been measured as a function of the magnetic field $B$, for three 
values of the driving frequency.  Experimental data are plotted as the 
sine of the loss angle $\varphi$ as function of the ratio 
$\Lambda/\mu_r\delta_f$, so as to verify the relation 
$\Lambda=\mu_r\delta_f(\Omega, B) \sin\varphi$ (straight line) 
resulting from Eq.(\ref{MSP}).  The limit $\sin\varphi=1/\sqrt{2}$ 
corresponds to the normal state or to the depinned vortex array.  The 
Campbell equation (\ref{CCC}) leads to much smaller loss angles: 
$\Lambda^2 = \frac{1}{2}\delta_f^2 \sin 2\varphi$ (dashed line).}
\label{sinePhi}
\end{figure}

\begin{figure}[!!!t]
\centerline{\epsfig{file=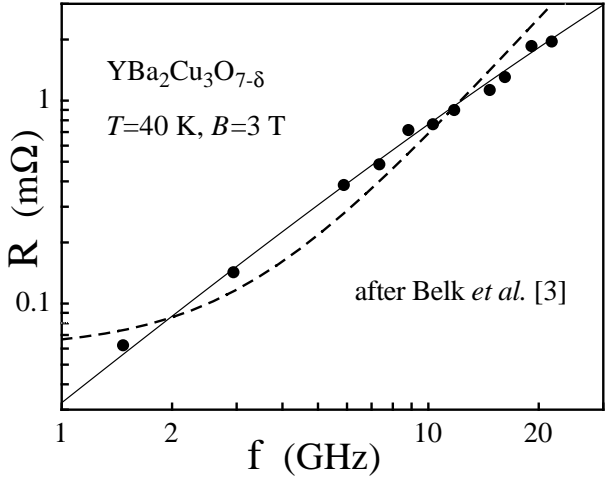, scale=0.45}}
\caption{The microwave surface resistance $R$ of an YBaCuO film 
(sample No.\thinspace 2 of Ref.~{\protect\cite{Belk96}}) vs frequency plotted in 
a log-log scale. ({\Large$\bullet$}) are experimental data taken from Fig.~6 of 
Ref.~{\protect\cite{Belk96}}. The full line is a fit using our 
Eq.~(\ref{MSP}), and taking $\lambda_S$ and $\rho_f$ (or $\lambda_f$) as 
two adjustable 
parameters (we find $\lambda_S=0.07\mu$m, $\rho_f=0.4\,\mu\Omega$cm, 
$\Omega_d/2\pi=100$ GHz). 
This fit is very close to the empirical power-law $R\sim f^{1.27}$  
proposed by the authors.
The dashed line shows the best fit obtained in 
Ref.~{\protect\cite{Belk96}} from the Coffey-Clem flux creep-model 
{\protect\cite{Coffey92}} }
\label{belk}
\end{figure}

As shown in Fig.~\ref{lambda}, the actual response is quite different:
after a low-frequency plateau,   
$\lambda^*\simeq\lambda'(0)$ ($\lambda''\simeq 0, \varphi\simeq0$),
the loss-angle increases with frequency, 
so that the ideal skin effect ($\lambda^*\simeq\lambda_f, \varphi=\pi/4$) 
is recovered beyond some depinning frequency $\Omega_d$;
this can be precisely defined as the mid-frequency for which 
$\varphi=\pi/8$. 
Note, in passing, that the observation of a linear response is not 
consistent with predictions of a naive critical state model: 
A critical-current density as small as $J_c\sim 10\,$A/cm$^2$ 
should restrict the penetration of fields $b_0\sim 1\,\mu$T to depths 
$\Lambda=b_0/\mu_0 J_c \lesssim 1\,\mu$m, which is much smaller than observed, 
and seeing that $\Lambda\propto b_0$, no linear regime could exist at all.  
The linear skin effect over depths $\Lambda\sim 100\,\mu$m was first 
reported by P. Alais and Y. Simon \cite{Alais67}, 
and then misinterpreted by considering the possibility of thermally 
activated vortex motion. 
Soon after, Campbell suggested that the linear signal was due to small reversible 
oscillations of vortices in their pinning potential wells \cite{Campbell69}: 
if a pinning restoring force $-nK\bbox{u}$ (per unit volume) is introduced 
in the equation of vortex motion, the propagation of the $k_1$-mode is greatly altered.  
At low frequencies, it becomes a non-dissipative evanescent mode decaying 
exponentially in the sample over a small depth $\lambda_C=(B\varphi_0/\mu_0 
K)^{\frac{1}{2}}\sim 1$--$100\,\mu$m.  
Here $\Omega_d=\rho_f K/B\varphi_0$ ($\delta_f(\Omega_d)=\lambda_C\sqrt{2}, 
\varphi=\pi/8$) \cite{Gittleman66}.  
Assuming $\mu_r\simeq 1$ and $\lambda_C\ll d$, 
the Campbell expression for $\lambda^*$ reads
\begin{equation}
-k_1^2 = \frac{1}{{\lambda^*}^2} = \frac{1}{\lambda_f^2} +
\frac{1}{\lambda_C^2}\quad.\qquad (d\gtrsim2\lambda_C)
\label{CCC}
\end{equation}
With $\lambda_f^2=i\delta_f^2/2$, Eq.~(\ref{CCC}) accounts for the 
low-frequency plateau and the related order of magnitude of 
$\Omega_d$.  Otherwise, no satisfactory fits of both 
$\lambda'(\Omega)$ and $\lambda''(\Omega)$ can be obtained from 
Eq.~(\ref{CCC}), as shown in Figs.~\ref{lambda} and \ref{sinePhi}.  In 
spite of recent attempts to improve the treatment of bulk pinning, the 
same difficulties are encountered in fitting $R(f)$ in YBaCuO 
\cite{Belk96}.  Note in this respect that the inclusion of thermal 
flux-creep effects \cite{Coffey92} may enhance the dissipation in an 
intermediate range of frequencies, as required (Fig.~\ref{belk}); it 
should be emphasized, however, that flux-creep models 
\cite{Coffey92,Alais67} predicts an unobserved divergence: 
$\lambda'=\lambda''\propto \Omega^{-\frac{1}{2}}$, as 
$\Omega\rightarrow 0$ \cite{Beek93}.

The model of the critical state based on the Mathieu-Simon (MS) 
continuum theory of the mixed state \cite{Mathieu88,Hocquet92}, has 
prompted us to an alternative interpretation.  We briefly recall the 
points of importance in the MS theory: \textbf{i)} each vortex line 
(unit vector $\bbox{\nu}$) must terminate normal to the surface 
($\bbox{\nu}\times\bbox{n}= 0$); whence the leading part of the 
boundary conditions (rough or smooth surface) in any problem of 
equilibrium or motion of vortices.  \textbf{ii)}~Vortex lines are not 
always field lines, so that the vortex field 
$\bbox{\omega}=n\varphi_0\bbox{\nu}$ and the mesoscopic field 
$\bbox{B}$ must be regarded as two locally independent variables.  The 
conjugate variable of $\bbox{\omega}$, 
~$\bbox{\varepsilon}=\varepsilon(\omega,T)\,\bbox{\nu}$, appears as a 
local line tension $\varphi_0\bbox{\varepsilon}$ (J/m) in the MS 
equation for vortex equilibrium or non-dissipative motion $\bbox{ 
J}_s+\mbox{curl}\bbox{\varepsilon}=0$.  \textbf{iii)}~The~classical 
picture of a local diamagnetism is misleading \cite{Vasseur97}.  A 
diamagnetic current, just like a subcritical transport current, is a 
true non-dissipative supercurrent $\bbox{J}_s$ 
($=-\mbox{curl}\bbox{\varepsilon}$) flowing near the surface 
over a small vortex-state penetration depth $\lambda_V$ 
($\lesssim\lambda_0$, the zero-field London depth).  Any deviation 
$\bbox{\omega}-\bbox{B}$ also heals beyond the depth $\lambda_V$, so 
that $\bbox{\omega}\equiv\bbox{B}$ in the bulk sample.  Although the 
mean magnetic-moment density of a perfect body turns out to be 
$-\bbox{\varepsilon}$, the quantity $-\bbox{\varepsilon}$ has not the 
primary physical meaning of a local magnetization, nor $\mu_r$, 
conveniently defined as the ratio $\omega/(\omega+\mu_0\varepsilon)$, 
that of a true local permeability \cite{Vasseur97}.

Let us return to the ac response of the slab in the standard geometry 
of Fig.~\ref{vrock}, and suppose that bulk motions are unrestrained (no bulk pinning). 
As pointed out by Sonin {\it et al.} \cite{Sonin92},
the distinction between $\bbox{\omega}$ and $\bbox{B}$ implies additional degrees of freedom, 
and a second $k_2$-mode appears besides the classical $k_1$-mode ($k_1=\pm i/\lambda_f$);  
this is a London-like non-dispersive evanescent mode,
which dies off over the depth $\lambda_V$ ($k_2=\pm i/\lambda_V$) \cite{Placais96}. 
Note that, in practice, $\lambda_V\ll\delta_f$, $\Lambda$ and $d$.  
In a pure $k_2$-mode vortex and field lines bend in opposite directions,
whereas they coincide in a pure $k_1$-mode. More explicitely 
\cite{Vasseur97,Placais96}, 
\begin{equation}
\nu_{1x}=\frac{b_1}{B}=\frac{u_1}{\lambda_f}\quad,\quad
\nu_{2x}= -\frac{ b_2}{B}~\frac{\mu_r}{1-\mu_r} = 
\frac{u_2}{\lambda_V}\quad,
\label{robe}\end{equation}
where $\nu_x=\partial u/\partial z$ ($\nu_{1x}=ik_1 u_{1x}$, 
$\nu_{2x}=ik_2 u_{2x}$) and $b/B$ measure the slopes of vortex and 
field lines respectively.  
Then the response $b(z)$ will be that combination of the modes, $b_1+b_2$, which 
satisfies the field continuity, $b_0=b_{10}+b_{20}$, as well as the 
correct boundary condition for vortex lines.  
The latter condition will determine the relative weights $\beta_1=b_{10}/b_0$ and 
$\beta_2=b_{20}/b_0$ of the modes ($\beta_1 + \beta_2=1$), and, 
therefore, the effective penetration depth $\lambda^* =
\beta_1\lambda_f + \beta_2\lambda_V \simeq \beta_1\lambda_f$ 
(since $\lambda_V\!\ll\!\delta_f$~and~$\Lambda$).  
For an ideal surface, the vortex boundary condition is clearly $\nu_x=0$ (point
\textbf{i)} above); then using Eq.~(\ref{robe}), we just recover the simple classical-diamagnetism result, that is 
$\beta_1=\mu_r$.  Now, the point is that the surface roughness may 
considerably change this boundary condition, so as to enhance the weight of the 
second mode (Fig.~\ref{vrock}a). 
Thus we argue that small effective skin depths at low frequencies 
should not result from restricting the penetration of the $k_1$-mode,
but from its amplitude being reduced due to the screening effect of the 
second mode.

According to the MS model 
\cite{Mathieu88,Hocquet92,Simon94,Placais94}, if the surface has 
irregularities on the scale of $a$, vortex lines can bend over a depth 
$\lambda_V$, making thus an angle $\alpha$ with the mean smoothed 
surface $z\!=\!d$.  On the average, and in any direction $x$, $\alpha$ 
should not exceed a critical value $\alpha_c\!\sim 1$--$10^\circ$ : 
$\langle\nu_x\rangle_{z=d}\leq\nu_{xc}\!=\sin\alpha_c\!\sim 
10^{-2}$--$10^{-1}$ \cite{Mathieu88,Hocquet92}.  As stated above, 
superficial non-dissipative supercurrents ($\bbox{J}_s 
=-\mbox{curl}\bbox{\varepsilon}$) can result from such distortions of 
the vortex array; integrating $-\mbox{curl}\bbox{\varepsilon}$, the 
net current density in the $y$-direction is $\mbox{\Large\it 
i}_y(\mbox{A/m}) = \langle\varepsilon_x\rangle= 
\varepsilon\langle\nu_x\rangle$, where $\varepsilon\simeq 
B(1-\mu_r)/\mu_r$.  A dc subcritical current can be regarded as a 
frozen pure $k_2$-mode in the limit $\Omega\rightarrow 0$.  To the 
maximum, $\mbox{\Large\it i}_y=\mbox{\Large\it 
i}_c=-\nu_{xc}B(1-\mu_r)/\mu_r$ is the surface critical-current 
density.  Starting from an equilibrium, where $\langle\nu_x\rangle=0$, 
a shift of the bulk vortex array is expected to induce a vortex 
curvature in the opposite direction, $\nu_x=f(u_{\text{bulk}})$ with 
$\nu_{xc}=f(u\!\sim\!a)$.  Perhaps such a \emph{vortex slippage} 
($\nu_x<0$ if $u>0$) is more intuitive when dealing with superfluid 
vortices in a rotating box (see Fig.~\ref{vrock}b).  In helium, bulk 
pinning does not exist, and only asperities of the walls can pin 
vortices \cite{Schwarz81}.  We are just extending this idea to 
collective motions of a vortex array along a rough surface of 
superconductors.

Linearizing $\nu_x=f(u_{\text{bulk}})$ for small displacements we can write 
$\nu_x\!=\!- u_{\text{bulk}}/l$, where $l$ is a real length characterizing the 
surface roughness: $l={\scriptsize\infty}$ would correspond to an ideal surface; in 
practice we expect that $l\sim a/\nu_c\sim 0.1-10\,\mu$m.  
As far as $\lambda_V\ll\delta_f$, this condition applies
quasistatically in the ac response by taking $u_{\text{bulk}}=u_{10}$,
so that the {\it vortex-slippage condition} reads
\begin{equation}
\nu_x=-\frac{u_{10}}{l}\qquad .
\label{slippage}\end{equation}  
Now, substituting  Eq.~(\ref{slippage}) for $\nu_x=0$ in the above calculation of 
$\beta_1$ and $\beta_2$, and considering the \emph{thick limit}, we obtain:
\begin{equation}
\frac{1}{\lambda^*} = \frac{1}{\mu_r\lambda_f} + 
\frac{1}{\lambda_S}\quad,\qquad \mbox{(\it thick~limit, $d\gtrsim2\delta_f$)}
\label{MSP}
\end{equation}
where $\lambda_S=l\mu_r/(1\!-\!\mu_r)\sim Ba/\mu_0\mbox{\Large\it 
i}_c$ is the real limit of $\lambda^*\simeq\beta_1\lambda_f$ as 
$\Omega\rightarrow0$ ($\lambda_f=(1+i)\delta_f/2$).  Note that setting $\mu_r=1$ from the beginning 
would lead wrongly to $\lambda^*=\lambda_f$, that is the ideal 
response.  While giving the same low and high-frequency limits as 
Eq.~(\ref{CCC}), $\lambda^*(0)=\lambda_S$ (real) and 
$\lambda^*({\scriptsize\infty})=\mu_r\lambda_f$ (depinning), 
Eq.~(\ref{MSP}) remarkably fits experimental data in the intermediate 
range $\Omega\sim\Omega_d$ (Figs.~\ref{lambda} and \ref{sinePhi}).  
According to Eq.~(\ref{MSP}), the graph $\lambda^*(\Omega)$ in the 
Argand diagram must be a quarter circle (Fig.~\ref{lambda}).  This 
universal behaviour should be easily tested in any case, irrespective 
of any adjustable or available parameters, only providing that the 
thick limit is achieved.

From data taken below the thick limit (not shown in the figures), we 
retain an important result.  When $2d$ is decreased under conditions 
where $\lambda_C\mbox{ or }\lambda_S\ll\delta_f$ 
$(\Omega\!\ll\!\Omega_d)$, we observe that the ac response becomes 
thickness-dependent as soon as $d\lesssim 2\delta_f$, as predicted: 
just substitute $\lambda_f\tanh(d/\lambda_f)$ for $\lambda_f$ in 
Eq.~(\ref{MSP}).  In particular, the loss angle is significantly 
smaller than stated by Eq.~(\ref{MSP}).  According to the one-mode 
Campbell model, size effects should arise for much thinner slabs such 
as $d\lesssim \lambda_C$.  Note in this respect that our ``pinning 
length'' $\lambda_S$, contrary to $\lambda_C$, does not represent an 
actual penetration depth.  The mere observation that size effects 
arise for $d\sim\delta_f$, not $d\sim\lambda_C$, attests that the 
bulk $k_1$-mode propagates freely, and reveals the need for a 
two-mode electrodynamics.
 
In conclusion, the MS model of the critical state completed by the 
vortex-slippage condition (\ref{slippage}) accounts quantitatively for 
the surface impedance of a variety of conventional samples, which all 
have standard critical-current densities ($\mbox{\Large\it i}_{\rm 
c}\sim 10$ A/cm): polycristalline lead-indium slabs and single-crystal 
slabs of pure niobium (Figs.~\ref{lambda} and \ref{sinePhi}).  For the 
application to the case of YBaCuO at $f\sim$ 10 GHz, our derivation of 
Eq.~(\ref{MSP}) has to be reexamined, especially because of the 
anisotropy and high-frequency correcting terms in the dispersion 
equation for the two-modes \cite{Placais96}.  Nevertheless, it is 
worth noting that Eq.~(\ref{MSP}) may account for the observed 
$f$-dependence of $R$ in YBaCuO from 1 to 20 GHz (Fig.~\ref{belk}) 
\cite{Belk96}.  These results support the argument, developed in 
previous works, \cite{Mathieu88,Hocquet92,Simon94,Placais94} that bulk 
pinning is absolutely ineffective in a large class of ``soft'' 
materials (devoid of strong bulk inhomogeneities).  Contrary to the 
common idea that any crystal defect may be a pinning center, we are 
led to the conclusion that a normally homogeneous sample in the mixed 
state rather imitates the behaviour of a superfluid vortex array 
enclosed in a rough box.

\end{document}